# HPS: A C++11 High Performance Serialization Library

*Junhao Li*
*Department of Physics, Cornell University*

Data serialization is a common and crucial component in high performance computing. In this paper, I present a C++11 based serialization library for performance critical systems. It provides an interface similar to Boost but is upto 1.5 times faster and beats several popular serialization libraries.

## Overview

Data serialization is a common and crucial component in high performance computing. It is a prerequisite for transferring data over the networks in distributed systems and for storing them on persistence storage devices. The bandwidths of the networks and the storage devices are the bottleneck of many distributed applications, such as MapReduce. Hence, a serialization framework that can produce compact serialized data efficiently can have game-changing impacts on these applications.

HPS is designed to and has achieved state of the art performance and beats well-known serialization libraries. Compared to Boost Serialization, HPS is upto 1.5 times faster and the serialized data from HPS are upto 40% smaller than those from Boost. Compared to Protocol Buffers, HPS is upto 4.5 times faster for certain types and the serialized data are upto 50% smaller.

The interface of HPS is similar to Boost, which means it works on standard containers and user defined structures directly without requiring users to deal with additional tightly coupled serialization classes. This increases code reusability and reduces cognitive efforts for the developers.

## Encoding Scheme

The encoding scheme of HPS is the root of its high performance and compact serialization. I heavily borrow the encoding scheme of Protocol Buffers while removing all the unnecessary fields in a performance critical setting. The resulting scheme contains only data and minimal structural information for recovering the original data unambiguously.

The integral types are encoded with base-128 varints. Signed integrals types use ZigZag encoding before the base-128 encoding. Hence, if we use HPS to serialize a number, such as -33, the serialized message takes only one byte. The floating point numbers are directly copied from the source memory location to the target. For iterable STL containers, HPS encodes their sizes first, followed by the elements. For custom classes, HPS looks for the serialize and parse methods in these classes.

In the following I give two encoding examples:

Case #1: A vector of integers

```
#include <cassert>
#include <iostream>
#include "../src/hps.h"

int main() {
  std::vector<unsigned> data({22, 333});

  std::string serialized = hps::to_string(data);
  auto parsed = hps::from_string<std::vector<int>>(serialized);

  assert(parsed == data);

  std::cout << "size (B): " << serialized.size() << std::endl;
  // size (B): 7
  // Serialized as (in hexadecimal):
  // 02 (number of elements)
  // 16 (first number)
  // cd (first byte of the second number in base-128 encoding)
  // 02 (second byte of the second number)

  return 0;
}
```

Case #2: Custom Class - A Quantum System

```cpp
#include <cassert>
#include <iostream>
#include "../src/hps.h"

class QuantumState {
 public:
  unsigned n_elecs;
  std::unordered_set<unsigned> orbs_from;
  std::unordered_set<unsigned> orbs_to;

  template <class B>
  void serialize(B& buf) const {
    buf << n_elecs << orbs_from << orbs_to;
  }

  template <class B>
  void parse(B& buf) {
    buf >> n_elecs >> orbs_from >> orbs_to;
  }
};

int main() {
  QuantumState qs;

  qs.n_elecs = 33;
  qs.orbs_from.insert({11, 22});
  qs.orbs_to.insert({44, 66});

  std::string serialized = hps::to_string(qs);

  std::cout << "size (B): " << serialized.size() << std::endl;
  // size (B): 7
  // HPS looks for the serialize<B> and parse<B> in QuantumState.
  // The first byte is n_elecs, then 3 bytes for each unordered_set.

  return 0;
}
```

Compared to Protocol Buffers, HPS does not store data types and field numbers. This reduces the size of the serialized message and also increases the performance. In the extreme case, if the leaf messages contain only non-repeating fields, the serialized message from Protocol Buffers will be 2 times larger than HPS.

## Implementation

There are two key classes in the implementation of HPS, the Serializer and the Buffer.

Buffer provides the read/write char(s) interfaces to the Serializer, and provides the "<<" and ">>" operators to the wrapper functions.

Serializer<DataType, BufferType> provides the logical serialize and parse methods of the given DataType. It uses the read/write char(s) methods of BufferType to push data to or pull data from the buffer. To make the codebase highly maintainable without sacrificing performance, we heavily use static polymorphism and SFINAE. For example, here is the Serializer specialization for the floating point numbers:

```
template <class T, class B>
class Serializer<T, B, typename std::enable_if<std::is_floating_point<T>::value, void>::type> {
 Public:
  static void serialize(const T& num, B& ob) {
    const char* num_ptr = reinterpret_cast<const char*>(&num);
    ob.write(num_ptr, sizeof(num));
  }

  static void parse(T& num, B& ib) {
    char* num_ptr = reinterpret_cast<char*>(&num);
    ib.read(num_ptr, sizeof(num));
  }
};
```

The wrapper functions call the Serializer to provide an easy to use interface to the users, for example:

```
std::ofstream out_file("data.log", std::ofstream::binary);
hps::to_stream(data, out_file);
```

```
std::ifstream in_file("data.log", std::ifstream::binary);
auto parsed = hps::from_stream<std::vector<int>>(in_file);
```

By default, Serializer<DataType, BufferType> will call the serialize<B> and parse<B> methods of the corresponding class. This provides a simple and loosely coupled way of extending HPS for custom classes (See the quantum system case above for example).

The complete implementation is hosted at https://github.com/jl2922/hps.

## Benchmark

The performance of HPS compared to other well-known C++ serializers for some most common data structures in high performance computing are as follows: (less is better)

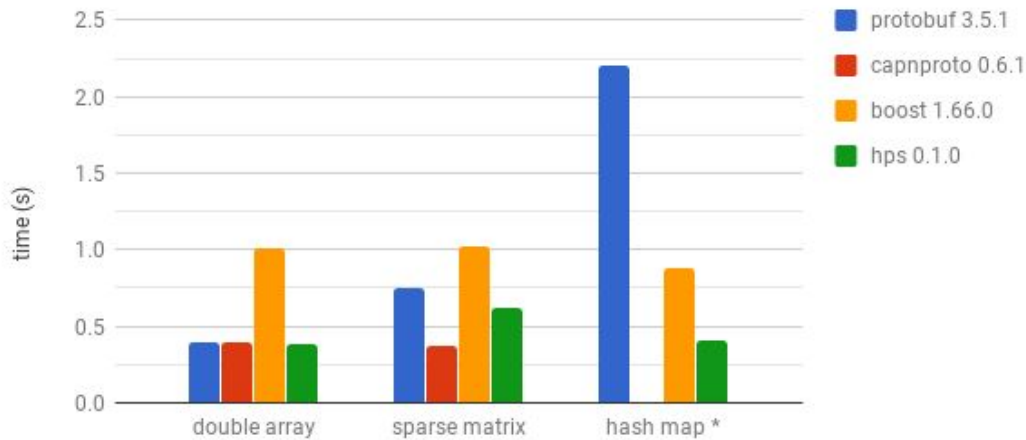

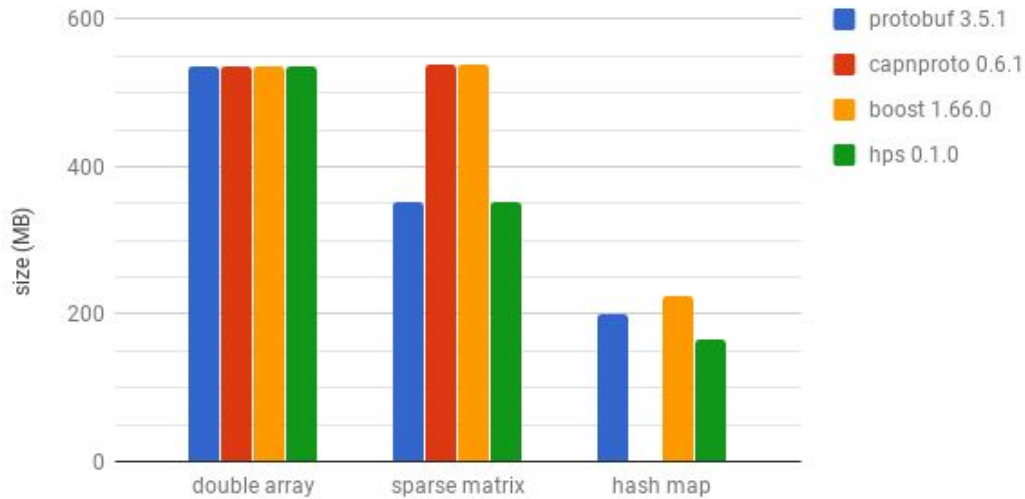

The sparse matrix is stored as a list of rows, each of which contains a list of 64-bit integers for the column indices and a list of doubles for the values. The hash map is a map from strings to doubles. Both HPS and Boost can serialize std::unordered_map directly, ProtoBuf uses its own Map type and CapnProto does not support hash map or similar types.

We can see that HPS is consistently and sometimes significantly faster than Boost and Protocol Buffers. It also beats Capnproto in all the cases except the sparse matrix. However, for the sparse matrix, although Capnproto is faster, the serialized message is also significantly larger, and for large distributed systems, minimizing the size of the data transferred is often more important than minimizing the local CPU time.

In addition to the traditional benchmarks for computational cost, we also provide the cognitive effort in terms of source lines of code for these test cases: (less is better)

| SLOC | double array | sparse matrix | hash map | fixed cost |
|---|---|---|---|---|
| protobuf | 12 | 23 | 12 | 17 |
| capnproto | 15 | 25 | - | 21 |

| boost | 13 | 20 | 13 | 13 |
| hps | 7 | 16 | 7 | 2 |

Note: fixed cost includes the estimated number of lines of commands needed for an experienced user to install the library, set the environment variables, extra lines of code needed in the Makefile, and various includes, etc.

Due to the header-only nature of HPS, it is the easiest one to set up, as we can see from the table above.

## Conclusion

In the paper, I present a high performance serialization library based on C++11. It is easy to use and beats the state of the art performance. Data serialization is important and may be the dominating factor of performance in many applications. HPS enables developers to serialize their structured data to compact binary formats efficiently so that their applications will incur less network traffic and disk operations and thus cost less and run faster.

## Acknowledgements

This work is supported by the U.S. National Science Foundation (NSF) grant ACI-1534965 and the Air Force Office of Scientific Research (AFOSR) grant FA9550-18-1-0095. We also thank professor Cyrus Umrigar for the helpful suggestions for the paper.